\documentclass[preprint2, usenatbib]{aastex}

\newcommand{\et}{\sl et al. \rm}

\slugcomment{SUBMITTED 2009-05-01; revised 2009-10-20, 2009-12-29, and 2010-02-01}

\shorttitle{Misaligned discs as AGN Obscurers}
\shortauthors{Lawrence and Elvis}

\begin{document}

%

\title{Misaligned Discs as Obscurers in Active Galaxies \\ \it\small Astrophysical Journal, 2010}

\author{Andy Lawrence\altaffilmark{1}}
\affil{Institute for Astronomy, University of Edinburgh, \\ Royal Observatory, Blackford Hill, Edinburgh EH9 3HJ, U.K.}

\and

\author{Martin Elvis}
\affil{Harvard-Smithsonian Center for Astrophysics, \\ 60 Garden Street, Cambridge, MA 02138}

\altaffiltext{1}{Visiting Scientist, Kavli Institute for Particle Astrophysics and Cosmology (KIPAC), Stanford University}

\begin{abstract}
We review critically the evidence concerning the fraction of Active Galactic Nuclei (AGN) which appear as Type 2 AGN, carefully distinguishing strict Type 2 AGN from both more lightly reddened Type 1 AGN, and from low excitation narrow line AGN, which may represent a different mode of activity. Low excitation AGN occur predominantly at low luminosities; after removing these, true Type 2 AGN represent 58$\pm5$\% of all AGN, and lightly reddened Type 1 AGN a further $\sim15\%$. Radio, IR, and volume-limited samples all agree in showing no change of Type 2 fraction with luminosity. X-ray samples do show a change with luminosity; we discuss possible reasons for this discrepancy. We test a very simple picture which produces this Type 2 fraction with minimal assumptions. In this picture, infall from large scales occurs in random directions, but must eventually align with the inner accretion flow, producing a severely warped disk on parsec scales. If the re-alignment is dominated by tilt, with minimal twist, a wide range of covering factors is predicted in individual objects, but with an expected mean fraction of Type 2 AGN of exactly 50\%. This ``tilted disc'' picture predicts reasonable alignment of observed nuclear structures on average, but with distinct misalignments in individual cases. Initial case studies of the few well resolved objects show that such misalignments are indeed present.

\end{abstract}

\keywords{galaxies:active -- galaxies:nuclei -- quasars: general -- accretion, accretion disks}

\section{Introduction}

The standard Unified Scheme for Active Galactic Nuclei (AGN) includes a central continuum source; a region somewhat further out emitting broad emission lines ; a dusty rotating ``torus'' on parsec scales; and gas emitting narrow emission lines on a scale of tens to hundreds of parsecs, ionised through the open cone defined by the torus edge \citep{le82, am85, kb86, kb88, up95}.  In this picture, Type 2 AGN are those seen sideways through the obscuring torus, so that one sees only the narrow lines and the IR emission produced by reprocessing of the continuum source by the torus. The torus needs to be geometrically thick ($H/R\sim 1$).  The shape and amplitude of the X-ray background (e.g. \cite{gilli07}), and the observed narrowness of emission line cones, appear to require that obscured (Type 2) AGN substantially outnumber unobscured (Type 1) AGN, by a factor $\sim 4$, so that the torus must subtend an angle of 65$^\circ$ seen from the central source (e.g. \cite{risaliti99}).

There is little doubt about the essence of the Unified Scheme concept - that most Type 2 AGN are the result of obscuration of the inner nucleus by some kind of flattened geometrically thick structure \citep{le82}. It is less clear that the specific rotating molecular torus model is correct. As recognised by \citet{kb88}, it is extremely difficult to maintain a cold rotating structure in a geometrically thick state, even if made of discrete clouds, or  puffed up by radiation pressure \citep{krolik07}.  Three alternative ways have been suggested to form geometrically thick obscuring structures. (i) A starburst disk in the host galaxy, kept turbulent and thick by supernova heating \citep{fabian98, wada02, thompson05, ballantyne08}. (ii) Dust bearing outflowing winds \citep{elvis02, elitzur06}; see also Fig. 8 of \citet{risaliti02}. (iii) Warped  disks \citep{phinney89, sanders89, greenhill03, nayakshin05, lawrence07}.  

In this paper we examine some specific predictions of the warped disk idea, common to all such models, regardless of their physical cause, under  the simplifying assumption of misaligned fuelling from completely random directions \citep{volonteri07}. We begin by looking critically at the issue of how many obscured AGN there really are (section \ref{f2}). Next we review what we know about parsec scale structures and gas flow in AGN (section \ref{structure}). We then test the predictions of the simplest misaligned disc models (section \ref{models}), and finally discuss the general implications of a misaligned disc interpretation (section \ref{discuss}).

\section{The Type 2 fraction, f$_2$}\label{f2}

Many authors use the terminology ``Type 2 AGN'' to mean any AGN showing signs of obscuration. In this paper we use the standard optical definition of a Type 2  AGN - that no broad emission lines or strong blue continuum are seen, while the narrow ($\sim$300--1000 km s$^{-1}$ FWHM) lines are prominent. However, where possible we also track the numbers of lightly reddened broad-line AGN, which show weak broad H$\alpha$ (see discussion below).  These can then be grouped with Type 2 or Type 1 AGN according to what hypothesis one wishes to test. Assessing what is and is not a Type 2 AGN can be difficult, for several reasons. 

(i) {\em Low quality spectra} can result in weak broad lines being missed. 

(ii)  Work reporting {\em X-ray surveys} often takes the existence of X-ray absorption as the definition of an ``obscured'' AGN, but in fact X-ray absorption can occur in the absence of optical extinction. Conversely, objects with very heavy optical extinction, such that the corresponding X-ray column is N$_H>$10$^{24}$cm$^{-2}$ (so called ``Compton Thick'' objects) may not be visible in X-ray surveys at all.  The generic term ``obscuration'' can hide some important distinctions; optical and X-ray classifications seem to disagree in around 20\% of cases \citep{tozzi06}.   

(iii) {\em Modest reddening} (i.e. A$_V\sim$1--3) can remove broad lines in the blue, resulting in what is often referred to in the literature  as objects of Type 1.8 or 1.9 \citep{ok76}. These are ``obscured'' AGN, but they are not Type 2 AGN, and typically have X-ray columns N$_H\sim$10$^{22}$ cm$^{-2}$, a hundred times smaller than true Type 2 AGN. To keep this distinction as clear as possible we refer to them below as ``reddened Type 1s'' or Type~1R, and track their numbers separately where these are reported. (The same terminology has been used recently by \cite{lacy07}.) The separation between Type 1R and Type 2 corresponds to approximately A$_V\sim$5, beyond which broad lines are extremely hard to detect in the optical; however this classification is especially sensitive to data quality, and to redshift. There may simply be a distribution of extinction values, but it has been argued that there are two physically distinct ``heavy'' and ``light'' obscuring regions, with the latter related to the host galaxy, and the former to the postulated nuclear ``torus'' (see \citet{mr95} and discussion below).  

(iv) {\em The nature of LINERs} which appear in many samples is still unclear. \citet{hopkins09} have recently made the case that at low luminosities in particular, the number of obscured AGN may have been substantially over-estimated, both because of data quality, and because of the prevalence of LINERs. 

Given these problems, it is particularly important to concentrate on statistics from reliable samples, and those that have minimum possible or well understood extinction bias. Table 1 summarises statistics from the various samples discussed below. Where authors report the number of Type~1R objects, we tabulate this; otherwise they are assumed to be included in the quoted number of Type~1 objects. However, some objects classified as Type~2 may eventually turn out to be Type~1R. We also tabulate, where known, the number of low-excitation AGN as distinct from classical high-excitation Type 2 AGN. (Table 1 refers to these as ``Type~L''). This fraction has an even higher degree of uncertainty.


\begin{deluxetable}{l l l c c c c c }
\tablecaption{AGN Type statistics}
\tablehead{
\colhead{Sample} & \colhead{Waveband} & \colhead{Ref} &  \colhead{Total} & \colhead{Type-L\tablenotemark{a}} & \colhead{Type-1\tablenotemark{b}} & \colhead{Type-1R} & \colhead{Type-2} \\
\colhead{}  & \colhead{}  &  \colhead{}   &  \colhead{}  & \colhead{f$_L$}   & \colhead{}   & \colhead{f$_{1R}$}  & \colhead{f$_2$} 
} 
\startdata
3C/6C/7C           & {\it radio} & (1)             &  323  & 107                       & 86         &                 & 130 \\
          &   &               &           &  0.33$\pm$0.04 &               &                 & 0.6$\pm$ 0.07  \\[5pt]
  
3CRR                & {\it radio/Mid-IR}  &(2)            & 42     & 17                           & 11        &                  & 14 \\
  &        &    &           &     0.4$\pm$0.12   &             &                  &0.56$\pm$0.19   \\[5pt]

Spitzer & {\it Mid-IR}  & (3) &            77  &        7        & 36         & 11           & 34 \\
             &          &            &          &  0.09$\pm$0.03               &               &0.16$\pm$ 0.05    & 0.49$\pm$0.1 \\[5pt]
  
IRAS-12$\mu$m & {\it Mid-IR}    & (4)           &  145  & 29                          & 53         &                  & 63 \\
        &    &             &          &   0.2$\pm$0.04    &               &                  & 0.54$\pm$0.08   \\[5pt]
  
IRAS-warm          & {\it Mid/Far-IR}  & (5)             & 226   & 5                            & 80          &                 & 141 \\
   &           &     &           &   0.02$\pm$0.01 &                &                 &0.64$\pm$0.07   \\[5pt]
  
RSA      & {\it optical}  & (6)        &   84   &             &  37         &  17           &  47\\
              &  &          &           &              &                 &  0.2$\pm$0.05 & 0.56$\pm$0.1  \\ [5pt]

Palomar-V1\tablenotemark{c}         & {\it optical} & (7)      &   218 & 167                        & 21         &                  & 30 \\
            &      &         &           &   0.77$\pm$0.08 &               &                 & 0.59$\pm$0.14   \\[5pt]
  
Palomar-V2\tablenotemark{d}         & &        &  218   & 122                       & 66         &                   & 30 \\
                      &         &              &             & 0.56$\pm$0.06 &                &                   & 0.31$\pm$0.07 \\[5pt]

Swift/BAT             & {\it Hard X-ray} & (8)  &           246  &       8        & 152         & 21            & 86 \\
  &               &  &        &  0.03$\pm$ 0.01              &               & 0.09$\pm$ 0.02 & 0.36$\pm$0.03 \\[5pt]
  
Swift/BAT   &        &  & {\it 332}   &     8          & 152         & 21            & {\it 172} \\
  corrected\tablenotemark{e}        &     &              &           &  0.02$\pm$ 0.01                &               & 0.06$\pm$0.02 & 0.53$\pm$0.03  \\[5pt]
\enddata
\tablenotetext{a}{``Type L'' is intended to refer to low excitation objects, but note that different authors define this class differently -- by weak OIII emission, by LINER-like spectrum, or by weak MIR emission.}
\tablenotetext{b}{The number of Type 1 objects includes any Type 1R objects.}
\tablenotetext{c}{Version-1 of the statistics from the Palomar sample places all LINERs in the Type L bin.}
\tablenotetext{d}{Version-2  places LINER-1s with the Type 1 AGN, and LINER-2s  in the Type L bin. }
\tablenotetext{e}{The number of Type 2 objects has been approximately corrected using the prescription of Risaliti \et 1999, that 50\% of all Type 2 AGN are Compton thick.}
\tablerefs{(1) Willott \et (2000); (2) Ogle \et (2006); (3) Lacy \et (2007); (4) Rush \et (1993); (5) De Grijp \et (1992); (6) Maiolino and Rieke (1995); (7) Ho, Filippenko and Sargent (1997); (8) Tueller \et (2009).}
\end{deluxetable}


\subsection{Radio selected AGN}  \label{radio}

In principle, selection of AGN by low-frequency radio emission is a safe method, being independent of both nuclear obscuration and beaming. (Of course this method only selects a subset of all AGN, those that are radio-loud.) \citet{lawrence91} showed that $f_2$ apparently changes systematically with radio luminosity in the 3CR sample, from $f_2\sim 0.5$ at L$_{178}$=10$^{29}$ W Hz$^{-1}$, to $f_2>$0.9 at L$_{178}$=10$^{25}$ W Hz$^{-1}$. However, several authors, while confirming the luminosity effect in other low-frequency radio samples, have pointed out that most of the non-quasar low-luminosity radio galaxies do not look like classical Type 2 AGN at all; they have low-excitation, and usually very weak, emission lines \citep{laing94, willott00, grimes04}.  A number of authors, using 3C, 6C, and 7C samples, have argued that such objects are not obscured Type~1 AGN at all, and represent a different mode of nuclear activity. They have optical cores dominated by synchrotron emission,  no sign of a mid-IR excess, and negligible nuclear X-ray emission, suggesting that there is no torus-like structure to extinguish the optical core, to reprocess hidden power, or to scatter hidden power into the line of sight \citep{chiaberge99, willott00, whysong04, ogle06, hardcastle06, hardcastle09, dicken09}.

In Table 1, we report statistics from the studies of \cite{willott00} (ref 1) and \citet{ogle06} (ref 2), as these authors explicitly correct for likely Type~L objects. They are consistent in showing $f_2\sim$0.6. The Type~L objects are nearly all at low radio power; \citet{willott00} report that after removing them there is no evidence for a luminosity dependence of f$_2$.

However, these studies did not explicitly report numbers of Type~1R objects. We know that this is significant issue; a careful study of 8 Narrow Line Radio Galaxies at Paschen $\alpha$ by \citet{hill96} converted 3 of these to reddened Broad Line Radio Galaxies, and several other such objects are known \citep{carleton84, goodrich92, economou95}.  Most objects in faint radio samples are at moderately high redshift so the available spectra are in the rest-frame blue; it is likely therefore that the value of $f_2\sim0.6$ is a significant overestimate.

\subsection{Mid-infrared selected AGN}  \label{IR}

A second reliable selection method is the mid-IR (5--20 $\mu$m), where the obscuring material re-radiates its absorbed energy,  where stellar contamination is minimal, and where emission by the cooler dust surrounding star forming regions is also relatively weak. Mid-IR emission should be close to being orientation independent, although there are some signs that the emitting region may be slightly optically thick at 12$\mu$m \citep{buchanan06}.  Table 1 shows numbers from three samples - a spectroscopic study of Spitzer 24$\mu$m sources (\citet{lacy07}, ref 3); the revised IRAS 12$\mu$m sample (\citet{rush93}, ref 4)  ; and the 60$\mu$m IRAS ``warm'' sample (\citet{degrijp92}, ref 5).  Overall, they show f$_2\sim$0.5--0.6.  The \citet{rush93} sample classifies many objects as LINERs, which as with the radio samples above, dominate at low luminosities. Removing these, there is no sign of a luminosity dependence of f$_2$ (see discussion in section \ref{lumdep}). \citet{lacy07} provide the number of Type~1R objects, which is 16\% of their sample.
 
\subsection{Nearby galaxy spectroscopic samples}  \label{specsamples}

A third reliable method is from complete spectroscopic surveys of nearby galaxies, which are volume limited down to quite small galaxy luminosities.  In Table 1 we show results from two studies; that of \citet{mr95} (ref 6), who examined spectral types for AGN for a sample constructed from the Revised Shapley Ames (RSA) catalogue; and the comprehensive Palomar spectroscopic study of nearby galaxies by \citet{hfs97} (HFS, ref 7).  \citet{mr95} exclude LINERs, but do provide a Type~1R number, which is large, at 20\% of the sample. This sample provides an interesting lesson in how one must be careful with terminology. The value derived from the \citet{mr95} sample, f$_2$=0.56$\pm$0.1, is consistent with Type~2 and Type~1 AGN occuring in equal numbers; but the sum of Type~2 and Type~1R AGN is 76\% of the sample, suggesting that ``obscured'' AGN outnumber ``unobscured'' AGN by 3 to 1.  Fairly small-seeming differences can lead to quite different conclusions.

The numbers from HFS illustrate another problem vividly. The HFS sample includes 157 objects classified as ``LINER'', but many of these have broad lines, and so are classified ``LINER-1''. Of the ``LINER-2''s, some may be obscured LINER-1s, and some may not. If we exclude all objects with a LINER classification completely (Palomar V1 in Table 1) then we get f$_2$=0.59, closely similar to other samples; if we include the LINER-1s as Type~1 (Palomar V2), then we get  f$_2$=0.31; this is obviously an underestimate, but we don't know how many of the LINER-2s to include. Perhaps the safest conclusion is to stay safely clear of the very lowest luminosity objects.

\subsection{SDSS AGN}  \label{sdss}

Several recent papers have considered the relative luminosity distributions of Type 1 and 2 AGN based on samples selected from the SDSS survey, but they have produced inconsistent results in the derived behaviour versus the luminosity in the [OIII] 5007 line.  At low luminosities, L(OIII)$\sim$10$^{33-34}W$, \citet{simpson05} finds  $f_2 \sim 0.8$, but \citet{hao05} and Netzer (2009) find  $f_2 \sim0.5$.  At high luminosities, L(OIII)$\sim$10$^{35-36}W$, \citet{simpson05} finds  $f_2 \sim 0.4$, \citet{hao05} finds  $f_2 \sim 0.25$, \citet{reyes08} find $f_2>$0.6, and Netzer (2009) finds $f_2 \sim 0.5$. (Note that the luminosity range in the figures of  \citet{hao05} differs from the other authors by a factor of 100, which we assume is an error. For Netzer (2009) we have estimated $f_2$ using the histograms shown in Fig 1 of that paper). Netzer (2009) stresses that there are difficult selection effects in using these data, especially at the low luminosity end; this is likely to be the origin of these discrepancies. A re-analysis of the SDSS data is underway to address these issues (Kewley  \et in preparation).

\subsection{Hard X-ray selected AGN}  \label{X-ray}

X-ray selected samples are not a reliable way to measure $f_2$, because of the bias produced by X-ray absorption. (The referee has asked us to stress that this should be considered the opinion of the authors.) However it is important to consider them, not least because of persistent findings that X-ray obscuration varies with luminosity \citep{le82, ueda03, hasinger08} - see section \ref{lumdep}. For broad-band flux limited surveys in soft or medium energy X-rays (e.g. 0.5 -- 2 keV or 2 -- 10 keV), with sources showing a wide range in $N_H$ and redshift, it is hard to correct for the effects of X-ray absorption. (This is for example clearly explained in \citet{dwelly06} : see their Fig. 15).  Samples selected in hard X-rays ($>$10keV; eg \citet{sazonov07, ajello08, tueller08}) are much better, because these should only be missing so called ``Compton-thick'' objects (i.e. those with N$_H>$2$\times$10$^{24}$cm$^{-2}$).  How many such objects appear in flux limited samples depends not just on the X-ray column density but also  on what fraction of the central luminosity is scattered back into the line of sight. Judging by the variation in L$_X$/L$_{OIII}$, the scattered fraction varies by an order of magnitude \citep{bassani99}, and there are examples of so called ``buried'' objects where the scattered fraction appears to be much less than 1\% \citep{bassani99, ueda07, sazonov08}. \citet{risaliti99} quantified the ``missing objects'' problem by examining X-ray spectra of a sample of optically selected AGN, finding that 50\% of all such optically defined Type 2 AGN  are Compton thick. \citet{sazonov07}, from spectra of hard X-ray sources found with INTEGRAL, find a somewhat lower number of thick sources - 10-15\% of all non-blazar AGN, i.e. perhaps a quarter of Type 2 AGN. However, their starting sample was hard X-ray based, so we do not know how many objects with very low scattered fractions will have failed to make it at all into the INTEGRAL catalogue. The \citet{risaliti99} result is therefore probably the best indication we have of how to correct for missing Compton thick objects.

Table 1 shows the results from the updated SWIFT/BAT hard X-ray catalogue (\citet{tueller08}; ref 8), who provide Type~1R as well as Type~1 and Type~2 numbers.  To these numbers we apply the Risaliti \et correction factor for likely missing Compton-thick objects (an extra Compton-thick object for every Compton-thin Type 2 AGN),  giving an estimate of f$_2$=0.53. This value is uncertain both because of the Compton-thick correction, and because of possible variation with luminosity - see section \ref{lumdep}.

\subsection{Luminosity dependence of f$_2$}  \label{lumdep}

It is important to know whether $f_2$ varies with luminosity or not. As noted in the subsections above, a luminosity effect in radio samples originally claimed by  \citet{lawrence91} was shown to be caused by the appearance at low luminosities of weak-lined low-excitation objects, which likely are not obscured Type 1 AGN,  and that the true Type 2 fraction does not change with luminosity \citep{willott00}. Likewise, when LINERs are discounted, the Rush et al 1993 IRAS MIR sample shows Type 1 and Type 2 luminosity functions which are closely matched throughout their luminosity range. Samples at very low luminosity (eg the Palomar sample of \citet{hfs97}) and at very high luminosity (eg the MIR/radio sample of \citet{ogle06}) agree in showing Type 1 and Type 2 AGN in approximately equal numbers. The situation is summarised in the lower panel of Fig. 1, where results from various surveys are superimposed on a common bolometric luminosity scale, estimated using the mean quasar spectral energy distribution from \citet{elvis94}. (This bolometric comparison is only approximate, but for comparisons over 5 decades will be adequate). From the radio, IR, and volume limited samples there is no evidence of variation in $f_2$ over 5 decades in luminosity.

The situation is strikingly different for X-ray samples. There are long standing claims that the prevalence of X-ray absorbed objects and/or Type 2 AGN changes with X-ray luminosity \citep{le82, ueda03, hasinger08}, but there have also been claims that this effect is in fact not present \citep{dwelly06, eckart06}. In the upper panel of Fig. 1 we compare the results from the hard X-ray sample of \citet{tueller08} with the result produced by \citet{hasinger08} from a compilation of medium energy X-ray surveys. After making the correction for missing Compton-thick objects described in the previous section, they agree very well, and disagree clearly with the behaviour shown by MIR and volume limited samples. \citet{gilli07} show that such a variation of obscured fraction with X-ray luminosity produces an excellent fit to the X-ray background and X-ray number counts, strongly suggesting that this is a real effect, not simply a problem with current studies. In principle, the X-ray background can also constrain the number of Compton-thick objects, but the discussion in \cite{gilli07} shows that this constraint is a very weak one, depending sensitively on the assumed scattered fraction.

However, objects known to be Compton thick are not completely absent from the Tueller et al sample - for example NGC 1068 is present. \citet{wang-jiang06} have argued that after removing known or suspected Compton-thick objects, the correlation of obscuration with luminosity is much less significant. It is striking that NGC 1068 is present with an apparent luminosity two orders of magnitude less than its likely true luminosity, usually assumed to be due to a small fraction of the nuclear light scattering back into the line of sight. We note that \citet{winter09} find that half the objects in the Swift/BAT sample require a complex spectral fit, for example partial covering. This raises the interesting possibility that many observed objects are partly covered with Compton thick material, and partly covered by material of intermediate thickness; such objects would be seen as apparently ``Compton thin'' but with a suppressed X-ray luminosity. This would produce an artifical correlation of obscured fraction with X-ray luminosity, and so could potentially produce the discrepancy observed in Fig. 1. Whether this produces a quantitatively correct explanation will be explored in a separate paper (Mayo \et in preparation.)

Another possibility for explaining the discrepancy in Fig. 1 is an effect due to evolution masquerading as a luminosity effect.  In X-ray samples, the question of the evolution of the observed obscured fraction has been controversial, but the conclusion seems to be a moderate observed change of the order $(1+z)^{0.4}$  (see discussion and analysis in \citet{hasinger08}). In the radio, \cite{willott00} find no significant evolution in the ratio of narrow-line to broad-line objects, but a mild evolution of the kind possibly seen in X-rays is hard to rule out. What about the IR ? In the lower pane of Fig. 1, most of the high luminosity points are from the IRAS sample of \cite{degrijp92}, with very few objects at $z>$0.1, and some from the Spitzer sample of \cite{lacy07}, with a median redshift of $z=0.6$, suggesting at most a 20\% effect due to evolution of the strength possibly seen in X-ray samples. This does not then seem to the explanation of the discrepancy seen in Fig. 1; however it would be highly desirable to undertake a fuller study of the luminosity and redshift effects in MIR selected samples. 

Another possible method of constraining the covering factor, at least in Type 1 AGN, is by asking what fraction of the bolometric luminosity emerges in the IR. The SED compilations of \citet{elvis94} and \citet{sanders89} show that this is typically 30\%, but with a large source-to-source variation. More recent studies of Type 1 AGN with Spitzer show that the ratio of mid-IR to optical luminosities is larger for lower luminosity AGN, and this has been interpreted as implying that the torus covering factor decreases from $\sim$ 90\% at L$_{bol}\sim$10$^{43}$ erg s$^{-1}$ $\sim$ to $\sim$ 20\% at L$_{bol}\sim$10$^{47}$ erg s$^{-1}$ \citep{maiolino07, treister08, hatziminaoglou09}. However the derivation of covering factors from L$_{IR}$ is highly model dependent, and there are likely selection effects such as starburst components being more dominant at low powers, and Type 1R AGN having artificially low apparent optical luminosities. The latest such study \citep{mor09} fits more detailed models to IR SEDs, including a clumpy torus and a starburst component. They find a mean torus covering factor of 0.27, and at best marginal evidence for an anti-correlation with luminosity.

There are some interesting and important effects which need further study, but IR, radio, and volume limited samples must be the most reliable. It therefore seems likely that the true value of $f_2$ does not change with luminosity, with as yet poorly understood selection effects causing an apparent trend in other studies.

\begin{figure}[h]

\includegraphics[scale=0.4, angle=0]{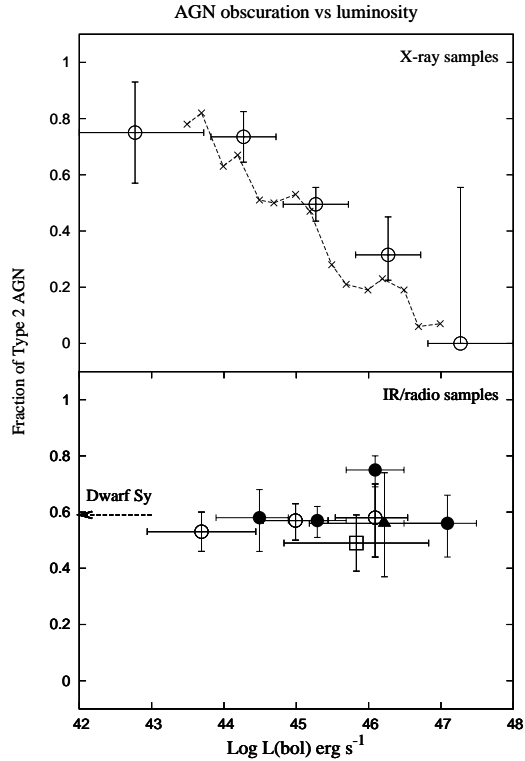}
\caption{\it Behaviour of $f_2$ with estimated bolometric luminosity. Upper pane, open symbols - data from Tueller et al 2009; crosses - Hasinger 2008.  Lower pane, filled circles - De Grijp et al 1992; open circles - Rush et al 1993;  open squares - Lacy et al 2007; filled triangles - Ogle et al 2006; dashed line - Ho et al 1997. Error bars are 68\% confidence  except for the lower limit at high X-ray luminosities, which shows a 90\% confidence interval.} \label{fig-1}
\end{figure}

\subsection{Conclusions on Type 2 fraction}  \label{f2-conc}

Table 1 summarises the results on $f2$ for different samples, and Fig. 1 shows the evidence on luminosity variation on $f_2$. As discussed in section \ref{lumdep}, the X-ray data show a luminosity dependence, but other selection methods do not, so it is reasonable to compare the non--X-ray samples. They are statistically consistent with a median value of $f_2$=0.56,  a weighted mean value of $f_2$=0.58$\pm$0.03, and a dispersion of $\sim$0.05. (Given the likely remaining systematic errors, the latter is a better indication of the uncertainty.) By contrast, the number of Type~L objects varies widely between samples, depending on both the selection method and the luminosity range covered by a sample. The fraction of Type~1R objects is poorly known, but is approximately 0.15$\pm$0.05.  All in all, a reasonable statement on current evidence is that out of every 100 AGN, 30 are relatively unobscured, 15 are lightly obscured, and 55 are heavily obscured. 

There may be a continuous distribution of degrees of obscuration. Alternatively there may be physically distinct light and heavy obscuring agents. If these agents are independent and occur randomly, then the number of lightly obscured objects may be twice as many as indicated above, as heavily obscured AGN may have additional light obscuration which goes unnoticed. In this case out of every 100 AGN, 30 AGN are unobscured, 55 are seen through a heavy obscurer, and 33 are seen through a light obscurer, including 18 out of the 55 heavily obscured AGN.  For example, \citet{mr95} noted that moderate obscuration, unlike heavy obscuration, correlates with host galaxy inclination, suggesting that there are two separate obscurers, one nuclear and one host-related. (This effect can also be seen in more recent faint X-ray samples, e.g. \citet{brand07}.  In this picture, we would interpret the AGN Type statistics as indicating that the nuclear obscurer has a covering factor of $\sim$ 55\%, and the host-related obscurer a covering factor of $\sim$ 33\%.

In the remainder of this paper, we consider only the heavy obscurer, which is almost certainly nuclear, and has a covering factor of just over 50\%.

\section{Nuclear structures and gas flow} \label{structure}

In section \ref{models} we look at predictions from a simple model of the obscuring region. In this section we examine evidence on the nature of the obscuring region in AGN, in order to constrain the ingredients of such a model.

\subsection{Location and size of obscurer} \label{location}

What do we know about the material constituting the geometrically thick obscurer ? We can estimate its distance if we assume that the dust observed radiating is part of the same structure responsible for the extinction - a good assumption, but not a certain one. The IR SED typically peaks around 10--20$\mu$m \citep{elvis94} implying a characteristic radiating temperature of $\sim 200K$. A minimum size is that of the blackbody producing this radiation, with R$_{BB} \sim$ 3 L$_{IR-44}^{1/2}$ T$_{200}^{-2}$ pc  where L$_{IR-44}$ is the infrared luminosity in units of 10$^{44}$ erg s$^{-1}$. A maximum size is set by the distance  at which an inefficient greybody dust grain is in equilibrium at T=200K when exposed to the full unattenuated heating power :  R$_{eq}$=37 L$_{44}^{1/2}$ T$_{200}^{-2.8}$ pc \citep{barvainis87}. 

Realistically the size is likely to be in between these two, set by model dependent radiative transfer effects, i.e. self-shielding. Model fits indicate that optical depths around $\tau\sim$3 are appropriate \citep{barvainis87}, so that the dust is at $\sim$1.8  L$_{44}^{1/2}$ T$_{200}^{-2.8}$ pc.  However, reverberation measurements in Seyfert galaxies at 2$\mu$m show delays on timescales of 10--100 days implying far smaller sizes \citep{suganuma06}, which must therefore relate to much hotter dust. At an assumed sublimation temperature of $T\sim$ 1500K,  R$_{eq}$=0.13 L$_{44}^{1/2}$ pc, equivalent to light travel time t=155 L$_{44}^{1/2}$ days. This is just consistent with the observed delays, but inconsistent with the SED peak.  AGN therefore contain dust at a wide range of temperatures, covering distances ranging from 0.1 pc to 10pc (at a fiducial luminosity of 10$^{44}$ erg s$^{-1}$). However, hot dust far outshines cool dust per unit area, at all wavelengths; for the SED to peak at 10--20$\mu$m, the bulk of the radiating area  must be at $\sim$ 1 pc. This dominance of the $\sim$ pc scale dust could be produced by geometric or self shielding effects, or both. If the luminosity is produced by accretion onto a black hole radiating at fraction $\epsilon$ of its Eddington luminosity, then  we can express this characteristic size scale as 
\begin{equation}
R/R_S \sim 1.5\times 10^6 e^{-\tau/3} (\epsilon/0.1) L_{44}^{-1/2} T_{200}^{-2.8}
\end{equation}
where R$_S$ is the Schwarzschild radius, $\epsilon$ is taken to be 0.1, and the dust optical depth is taken to be $\tau$=3.

Current mid-infrared interferometric measurements, with $\sim$10mas resolution, can just resolve such scales in the very nearest AGN, and confirm parsec size scales directly in NGC~1068 and Circinus \citep{jaffe04, tristram07, raban09}.  Cold gas is also detected via maser emission, and in a handful of cases has been beautifully resolved by VLBI measurements - in NGC~1068 \citep{gg97, gallimore04}, in NGC~4258 \citep{greenhill95, miyoshi95, herrnstein05, humphreys08}, and in Circinus \citep{greenhill03}. These observations are consistent with thin Keplerian disks at variously convincing levels, from NGC4258 (convincing) to  NGC1068 (possible). The discs are warped but not very strongly so; we need to be careful of the selection effects that make masers visible. Some discussions (e.g. that in \citet{raban09}) assume that the masers are showing the outer edge of the accretion disc, with the puffed-up  ``torus'' lying just outside this, but those working on the maser observations themselves have suggested that the observed warped disc itself produces the obscuration \citep{greenhill03, herrnstein05, fruscione05}.  NGC~1068 also shows a cold warped disk structure in CO emission \citep{schinnerer00} but this is on considerably larger scales, 50--100 pc. Finally, an intriguing new observation of molecular H$_2$ in NGC~1068 has been made by \citet{sanchez09} using Adaptive Optics and IR integral field spectroscopy, on  a scale of 10pc. This shows neither a disk-like structure, nor a ``torus'', but two linear streamers apparently pointing to and moving towards the nucleus. 

\subsection{Orientation of obscurer} \label{orientation}

What do we know about the orientation of the cold gas with respect to the accretion disk and black hole spin axis ? In NGC~1068, NGC~4258, and Circinus, the parsec scale maser and MIR material are oriented at least roughly perpendicularly to the radio axis  \citep{greenhill03, herrnstein05, raban09}. On the other hand, the radio axis seems to be randomly oriented with respect to the host galaxy in Seyfert galaxies \citep{uw84, clarke98}. The situation with radio galaxies has been more confused. \citet{schmitt02} describe the debate on this issue, but conclude that the radio axis  is randomly oriented with respect to the kpc scale dust disks often found in the elliptical host galaxies.  For both Seyferts and radio galaxies however, the data are also consistent with being random over a large polar cap \citep{kinney00, schmitt02}. A recent large statistical study of radio sources coincident with SDSS galaxies \citep{battye09} found clear evidence that the radio major axis is ``biased towards'' being aligned with the optical minor axis in early type galaxies, but this is seen as a $\sim$ 15\% excess of aligned galaxies, consistent with the distribution being random to first order.

So it seems that the AGN itself has a well defined axis on parsec scales and below, but that this axis is, to first approximation, unconnected with the kpc scale structure it finds itself in. 

\subsection{Gas flow in nuclear region} \label{gasflow}

What motion do we expect of gas reaching parsec scales ? Central molecular discs in galaxies are turbulent and have scale heights of the order 100 pc (e.g. \citet{scoville93}), one to two orders of magnitude larger than the gravitational sphere of influence of the black hole. Velocity mapping of H$_2$ emission lines in AGN on 30 pc scales using adaptive optics \citep{hicks09} shows net rotation but with a velocity dispersion of a similar size - i.e. very thick discs. Both in our own Galactic Centre and in nearby AGN, this observed molecular material is almost certainly in a large number of small dense clumps, so that the covering factor to the nucleus is of the order 1\% (see discussion in section 4 of \citet{hicks09}). This 30pc scale thick disc is therefore {\em not} the ``torus''; the nuclear obscuration must typically happen closer in. 

On 10 parsec scales, the movement of clumps is likely to be dominated by turbulence rather than rotation, so that motion will be close to isotropic. A tiny fraction of clumps (those on quasi-radial orbits) would be captured into the black hole sphere of influence. This seems to be just the picture shown by recent simulations, which find accretion rates (of captured material) that are highly variable on timescales of 10$^5$ yr, with fluctuations of two orders of magnitude \citep{wada02, wada04, escala07, nayakshin07}. This seems to be borne out by observations of NGC~1068 \citep{sanchez09}, which show molecular streamers on quasi-radial orbits unconnected with the central disc orientation, on a scale of 5-20pc.  

A good working hypothesis is therefore that AGN are fuelled by material arriving from 10--100pc scales, in discrete events coming from random directions. Each event may create a well defined accretion disc, but the ensemble of events over a long timescle has no net direction. 

\subsection{Nuclear warped discs} \label{warps}

The combination of a well-defined central accretion axis and material arriving from random directions on parsec scales suggests that a re-alignment of material must occur in between, producing extreme warping.

Warping of accretion discs has been discussed a number of times in the AGN literature. \citet{pringle96} showed that a radiation pressure instability can induce extreme warping in self-illuminated discs. \citet{nayakshin05} showed that in the transition region between the kpc scale disc and the sphere of influence of the black hole, the massive outer disk produces a torque on the inner accretion disc, inducing strong warping on a few orbital timescales. 

In contrast to these processes for {\em inducing} a warp, viscosity will try to keep a disc aligned over time \citep{bp75, pringle92}, although there has been dispute over whether the black hole aligns the disc or vice versa \citep{king05, volonteri07}.  The warp radius, where the inner aligned planar disc meets the outer misaligned disk, could be the location of the AGN obscuring structure, although current models place this at $\sim$10$^4$R$_S$, rather less than the size scale of the AGN obscurer discussed in the previous section ($\sim$10$^6$R$_S$, see eqn-1).

The true dynamics, and the warp induction or reduction mechanism, are beyond the scope of this paper. However, the hypothesis of random fuelling direction alone makes strong predictions which are generic and testable. We now look at these predictions.

\section{Tests of misaligned disc models} \label{models}

In this section we look at generic tests of the hypothesis that incoming material is randomly misaligned, and somehow adjusts - first looking at the distribution of covering factors and predicted value of $f_2$, and then at possible misalignment effects.  We need to consider both the {\em tilt} of the warp - how the misalignment between the spin axes of the central engine and each annulus varies with radius, $\theta$(R); and the {\em twist} of the warp - how the azimuth of the line of nodes changes with radius, $\phi$(R). Note that the twist refers to the shape of the structure at an instant in time, not to the precession - in principle each annulus could be precessing but locked together. We can distinguish two extreme assumptions : 2$\pi$ twist, and zero twist.

\subsection {Covering factor for fully twisted discs}

If the twist is 2$\pi$ or more, a complete equatorial wall is formed; a symmetrical structure that would look like the traditional ``torus''. If the initial misalignment is $\theta$, then the covering factor is $C=\sin\theta$. The probability of a given $\theta$ is  $dP=(\sin\theta/2) d\theta$. We then find the distribution of covering factors to be 

$$dN=\frac{1}{2} \frac{{C}}{\sqrt{1-C^2}}dC$$ 

Objects with large misalignments are more likely to be observed as Type 2 AGN;   the distribution of covering factors for only those objects which appear to be Type 2 is dN$_2$=C$\times$dN(C), and the distribution for Type 1s is dN$_1$=(1-C)$\times$dN(C). The observed Type 2 fraction is then $f_2 = \int$dN$_2$(C).  

Fig 2 shows the distribution of C for objects with  $\theta<\pi/2$. Note that {\em all} objects with $\theta>\pi/2$ (i.e. the predicted 50\% where the incoming flow is counter-rotating) are completely obscured from all directions. Including all the counter-rotating cases (f$_2$=0.89), this model is in strong disagreement with the observed value of $f_2\sim$0.55. Possibly the completely obscured counter-rotating cases would not even be recognised as AGN; they would perhaps be seen as ultraluminous IR galaxies, as suggested by many authors. However even ignoring the counter-rotating cases, this model makes too many obscured objects ($f_2\sim$0.8).  

We can therefore reject the hypothesis of twisted warped discs with random incoming directions.

\begin{figure}[h]

\includegraphics[scale=0.3, angle=270]{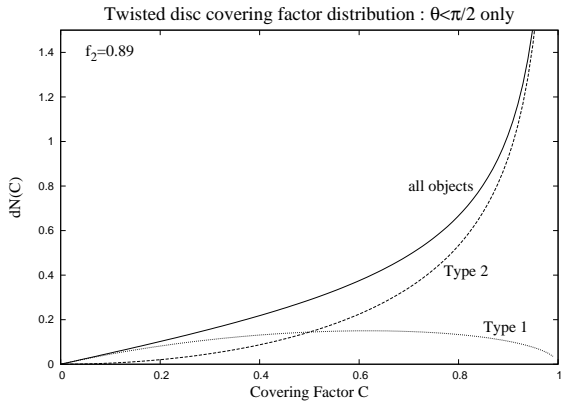}
\caption{\it Covering factor distributions for a misaligned disc with random incoming rotation directions, where the line of nodes rotates at least one cycle during re-alignment.} \label{fig-2}
\end{figure}

\subsection{Covering factor for tilt only discs}  \label{tilted}

For tilted discs with no twist, the covered sky is determined by the outermost annulus; this, together with interior annuli, defines a lune on the sky, so that the covering factor is  C=$\theta/\pi$ and the distribution of covering factors is 

$$dN=\frac{1}{2}\sin{\pi C} dC$$

Fig. 3 shows the covering factor distributions. The result is that $f_2=0.5$ exactly; a value remarkably close to the observed value, with minimal assumptions.  Note that the covering factor distributions for objects seen as Type 1 and Type 2 are not the same. The typical covering factor for Type 1 AGN is $\sim0.35$, close to the fraction of bolometric luminosity which emerges in the IR \citep{sanders89,elvis94}, and to the typical torus covering factor found in recent model fits (Mor et al 2009).

\begin{figure}[h]

\includegraphics[scale=0.3, angle=270]{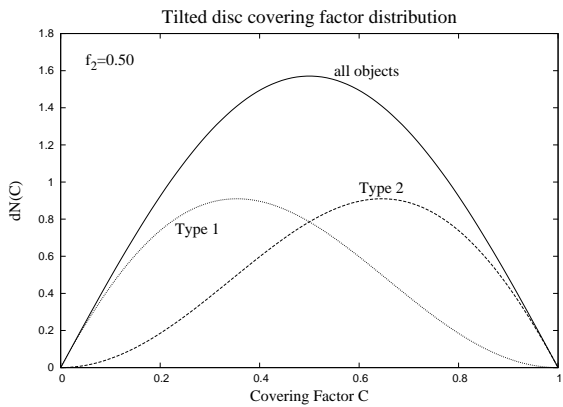}
\caption{\it  Covering factor distributions for a misaligned disc with random incoming rotation directions, where the line of nodes has no rotation during re-alignment. } \label{fig-3}
\end{figure}

\begin{figure}[h]

\includegraphics[scale=0.35, angle=0]{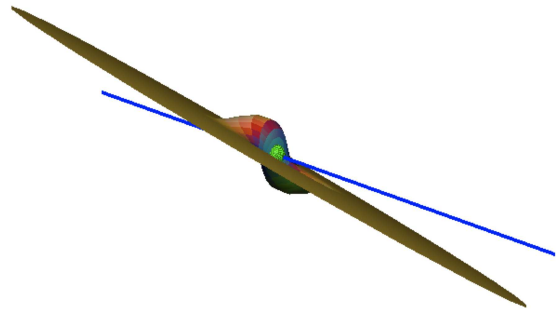}
\caption{\it  Illustration of what a tilted disc could look like seen almost edge-on, and close to the line of nodes.  This example has an initial misalignment of $\theta$=80$^\circ$. The green ball represents the central source, and the blue line the radio jet.
} \label{fig-4}
\end{figure}

\subsection{Intermediate cases} \label{intermediates}

Obviously more subtle parameterisations are possible; for example, a small amount of twist will increase the mean covering factor.  Likewise fully twisted discs could be viable as a hypothesis if we give up the assumption of random misalignment. For example, if we assume that the probability of a given misalignment angle $\theta$ is smaller at large $\theta$ compared to a random distribution, e.g.  tapered by a factor $\cos^2\theta$, so that $dP=3\sin\theta\cos^2\theta d\theta$, then we find that the Type 2 fraction is $f_2$=0.59. However, a physical basis for such intermediate hypotheses would be needed before knowing what to test; we leave this to future work, and instead look at one more area of generic tests.

\subsection{Predicted misalignment effects} \label{misalignments}

We have seen above that randomly misaligned discs with no twist pass the covering factor test. However, such discs are not azimuthally symmetric. Although warped disk obscurers can produce shadow cones  \citep{greenhill03}, and on average these will align with the nuclear axis, such a model predicts that distinct misalignments of nuclear structures will frequently be seen. An illustrative example is shown in Fig 4. In this section we test for such misalignments in a few exceptionally well observed  individual cases, each one the brightest and nearest of its class - Type 1 Seyfert, Type 2 Seyfert, FRII radio galaxy. We find that interesting alignment anomalies are indeed present.

{\bf NGC~4151}. Fig. 5 is a sketch indicating the structures seen in NGC~4151. As a Type 1 AGN, in the usual ``torus'' scheme, NGC~4151 should not show emission line cones at all, but it  does \citep{hutchings98}. This situation is usually, though vaguely, explained as
due to our observing at an angle grazing the torus. However, in the tilted disc picture, this combination is easy to achieve over a significant range of azimuths. The radio and emission line structures are roughly co-aligned, but the jet does not bi-sect the emission line structure \citep{mundell03}. Rather, it lies along one edge. There is a marginally resolved structure in molecular $H_2$ (shown in blue in Fig. 5) which has been claimed as representing the ``torus'' in NGC~4151 \citep{fernandez99} but this has a minor axis which differs from the radio axis by $\sim$80$^\circ$. 

\begin{figure}[h]

\includegraphics[scale=0.38, angle=0]{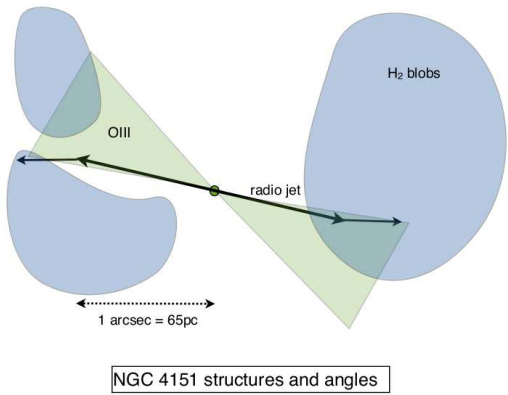}
\caption{\it Sketch of structures seen in NGC~4151. The radio and emission line structures (black line and green region respectively) are taken from data and figures in \citet{mundell03}; the H$_2$ structures (blue regions) are taken from data and figures in \citet{fernandez99}. 
} \label{fig-5}
\end{figure}

{\bf NGC~1068.}  This Type 2 object now has a rich variety of data. Fig 6 is a development of the illustration recently shown by \citet{raban09}. (Note that this figure is not on a linear scale, and mixes structures on sub-pc and tens of pcs scales).  The inner radio jet is orthogonal to the  ``hidden BLR'' polarisation angle to within 5$^\circ$, suggesting that both of these correctly indicate the (presumed) black hole spin axis. On sub-pc  to 10 pc scales, elongated structures are  seen in all four of maser emission, mid-IR emission, radio continuum, and H$_2$ emission, but the orientation changes systematically with radius. The minor axis is misaligned with the jet axis by $\sim$ 20$^\circ$ on a scale of $\sim$ 0.1--0.4pc (inner maser disc),  by  $\sim$ 40$^\circ$ on $\sim$ 0.4--1pc scale (outer maser disc and mid-IR structure), and by  $\sim$ 70$^\circ$ on 1--10pc scales (H$_2$ streamers).  The region as a whole seems to show incoming material approximately N-S and a final major axis which is approximately E-W.  Meanwhile the radio jet changes direction at ``knot C'', at a distance of $\sim$ 20pc from the nucleus (see \citet{gallimore04}). The outer jet is roughly aligned with the extended narrow emission-line region on $\sim$50--100 pc scale, and both are misaligned with the inner radio jet by $\sim$ 20$^\circ$

\begin{figure}[h]

\includegraphics[scale=0.38, angle=0]{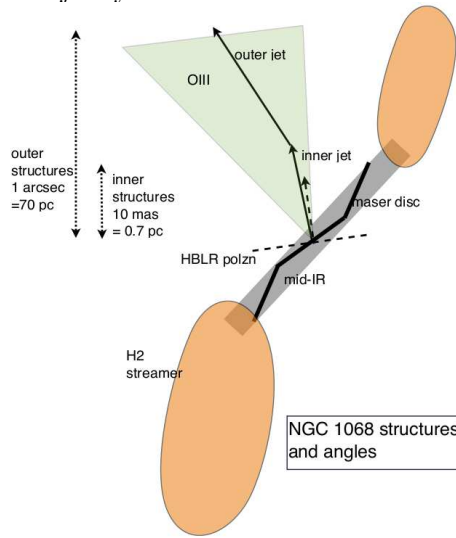}
\caption{\it  Sketch of structures seen in NGC~1068. (Note that this figure is not on a linear scale, and mixes structures on sub-pc and tens of pcs scales). The radio continuum and maser structures (black lines and dotted arrow) are taken from data and figures in \citet{gallimore04} and references therein; the mid-IR structures (grey region)  are from \citet{raban09}; the emission line structure (green region) are from \citet{evans91}; the BLR polarization angle (dotted line) is from \citet{am85}; and the H$_2$ structures (orange regions) are from \citet{sanchez09}.
} \label{fig-6}
\end{figure}

{\bf Cygnus A.} In near-IR light, this object shows an impressive X-shaped structure, suggestive of a hollow bi-cone, very well aligned with the radio axis, which is presumably made by scattered light \citep{tadhunter99}. However, IR polarisation imaging shows that back reflection seems to occur only along one ``wall'' of that bi-cone \citep{tadhunter00}. 

In conclusion, in well studied sources, there is clear evidence for misalignments, as expected for tilted disc obscurers. More detailed testing will require physical models. For example, scattered light ``cones'' will not be symmetric, but their surface brightness profiles will for example depend on how the density of scatterers varies with elevation.

\section{Discussion} \label{discuss}

Here we briefly mention some further implications of misaligned discs, and especially of simple tilted discs, as AGN obscurers. 

(i) In the misaligned disc picture, Type 1 and Type 2 AGN are still drawn from a single population, but they are not identical : Type 2 AGN are on average those objects with larger misalignments, and so larger covering factors. This potentially explains several interesting observations. 
\begin{itemize}
\item {The fraction of the bolometric luminosity of Type 1 AGN which emerges in the IR is smaller than the fraction of Type 2 AGN \citep{sanders89,elvis94,mor09}.}
\item{Type 2 AGN on average have more prominent dust lanes \citep{malkan98, hunt04}.}
\item{Type 2 AGN have weaker narrow-line emission for a given radio power \citep{whittle85, jb90,  lawrence91, grimes04}; this could be because they have narrower opening angles, as they would if more tilted or twisted.}
\end{itemize}

(ii) For both the twisted disc and tilted disc pictures, 50\% of cases would result in any polar outflow (e.g. radio jet) colliding with the disc. The implications of this would depend on the mass of the disc compared to the momentum of the outflow. If the disc stops the outflow, then this predicts that not {\em all} Seyfert galaxies will have radio jets. If the outflow breaks through the disc, then the twisted disc model could be revived; incoming material would usually cover most of the sky, but the outflow produces  a broken shell of obscuration. (This was  suggested by \citet{lawrence91}). Jet-ISM collisions have of course often been discussed before, for example as an explanation for Compact Steep Spectrum (CSS) radio sources (see review by \citet{odea98} and recent results by \citet{guainazzi06}). Jet-ISM interaction was proposed as the {\em cause} of the severe warp in 3C 449 \citep{tremblay06}, and has often been discussed in the context of the formation of the Narrow Line Region gas \citep{taylor92,bicknell98,dopita02}.

(iii) We assumed for simplicity that incoming material has no effect on the spin of the central engine, but of course it will have, depending on how large each accretion event is. However, if incoming events arrive at random, then there will be no net spin-up. The fractional angular momentum change produced by each event (assumed to be of equal mass)  will decrease with time as the black hole grows, so that the spin of the black hole does a  random walk, gradually converging on a spin value which is well below maximal spin \citep{volonteri07}. It has been argued that the difference between radio loud and quiet AGN is in the spin of the black hole \citep{wilsoncolbert95}, and  \citet{Sikora07} specifically argue that while radio quiet AGN are indeed fed by many small accretion events from random directions, resulting in a low final spin, radio-loud AGN are fed by major mergers with a well determined angular momentum.  If we can measure the distribution of tilt angles in AGN, then we may have an extra quantitative handle on these questions.

\section{Conclusions} \label{conc}

If one considers only reliable samples (i.e. those selected in the low frequency radio, in the mid-IR, or volume limited samples), removes low excitation objects, and includes lightly reddened objects with the Type 1 AGN, then the fraction of true, heavily obscured, Type 2 AGN is approximately 55$\pm5$\%. This value is independent of luminosity in IR, radio, and volume limited samples, but apparently changes with observed X-ray luminosity. The fraction of lightly reddened (but still recogniseably broad-line) AGN is 15-30\%, depending on whether this light obscuration occurs independently of the heavy obscuration. The heavy obscuration region is very likely the same as the parsec-scale IR emitting region. 

Inflow from kpc scales to this parsec scale is likely to be in random directions, misaligned with the inner accretion flow, and thus should lead to severely warped structures. Random incoming discs with with complete twist produce too many obscured objects, but warps with no twist produce the correct fraction of Type 2 AGN. As, in this simple model, the mean tilt in Type 2 AGN is larger than
in Type 1 AGN, the relative strength of the IR and weakness of [OIII]
are accounted for. This ``tilted disc'' picture also predicts significant misalignments of nuclear structures. Case studies of well resolved objects show that such misalignments do indeed occur.


\section{Acknowledgements}  \label{acknow}

This paper was written while AL was on sabbatical at KIPAC in Stanford, especially during a collaborative visit by ME. We would like to thank the staff of KIPAC for their hospitality and for the stimulating atmosphere.  At various stages we have had very useful discussions with a number of colleagues that have helped to shape this paper : we would particularly like to thank Sterle Phinney, Andrew King, Reinhard Genzel, Ski Antonucci, Jean-Pierre Lasota, Marek Sikora, and Lukasz Stawarz. Responding to the robust criticisms of an anonymous referee also helped us to improve the paper considerably. 


\bibliographystyle{apj}
\bibliography{lawrence-elvis-warp-arxiv}

\end{document}